\documentclass[11pt]{article}
\pdfoutput=1
\usepackage{amsmath}
\usepackage{amssymb}
\usepackage{mathrsfs}
\usepackage{bbm}
\usepackage{cancel}
\usepackage{mathtools}
\usepackage{slashed}
\usepackage[numbers,sort&compress]{natbib}

\usepackage{geometry}
\usepackage{float} 
\usepackage{tabulary} 
\usepackage{soul} 
\usepackage{subfigure}
\usepackage{graphicx}
\usepackage[section]{placeins} 
\usepackage{hyperref}
\usepackage[normalem]{ulem}
\usepackage[dvipsnames]{xcolor}

\usepackage{cleveref}
\crefname{equation}{Eq.}{Eqs.}
\crefname{figure}{Fig.}{Figs.}
\crefname{table}{Table}{Tables}
\crefname{Section}{Section}{Sections}
\allowdisplaybreaks

\def \et{{\it{~et al.,~}}}

\begin{document}

\title{\textbf{ ~~~~~~~~~~~~~~~~~~~~~~~~~~~~~~~~~~~~~~~~~~~~~~~~~~~~~~~~~~
{\scriptsize{MPP-2015-274}}\\~\\
ATLAS Diboson Excess from Stueckelberg Mechanism}}

\author{
Wan-Zhe~Feng,\footnote{Email: vicf@mpp.mpg.de}\\
\textit{Max--Planck--Institut f\"ur Physik (Werner--Heisenberg--Institut), 80805 M\"unchen, Germany}\\~\\
Zuowei Liu,\footnote{Email: zuoweiliu@nju.edu.cn}\\
\textit{Department of Physics, Nanjing University, Nanjing, 210093, China}\\
\textit{Institute of Modern Physics and Center for High Energy Physics,}\\
\textit{Tsinghua University, Beijing, 100084, China}\\~\\
Pran Nath\footnote{Email: nath@neu.edu}\\
\textit{Department of Physics, Northeastern University, Boston, MA 02115-5000, USA}
}

\date{}
\maketitle

\begin{abstract}
  We discuss the diboson excess seen by the ATLAS collaboration around 2 TeV
  in the LHC run~I at $\sqrt{s}=8$ TeV.
  We explore the possibility that such an excess can arise from a $Z'$ boson which acquires mass through
  a $U(1)_X$ Stueckelberg extension.
  The corresponding $Z'$ gauge boson is leptophobic with a mass of around 2 TeV  and has interactions with $SU(2)_L$ Yang-Mills fields and gauge fields of the hypercharge.
  The analysis predicts $Z'$ decays into $WW$ and $ZZ$ as well as into $Z\gamma$. Further three-body as well
  as four-body decays of the $Z'$ such as $WWZ$, $WW\gamma$, $WWZZ$ etc are predicted.
  In  the analysis we use the helicity formalism which allows us to exhibit
  {the helicity structure of the $Z'$ decay processes in an  transparent manner.
  In particular, we are able to show the set of vanishing helicity amplitudes
  in the decay of the massive $Z'$ into two vector bosons due to angular momentum conservation
  with a special choice of the reference momenta.}
  The residual set of non-vanishing helicity amplitudes are identified.
  The parameter space of the model compatible with the diboson excess seen by the
  ATLAS experiment at $\sqrt s=8$ TeV is exhibited.   Estimate of the  diboson excess expected at $\sqrt s= 13$ TeV with 20 fb$^{-1}$ of integrated  luminosity at LHC run~II is also given. It is shown that the $WW$, $ZZ$ and $Z\gamma$ modes
  are predicted to be in the  approximate ratio
  $1:\cos^2\theta_W (1+ \alpha \tan^2\theta_W)^2/2: (1-\alpha)^2\sin^2\theta_W/2$
  where $\alpha$ is the strength of the coupling of  $Z'$ with the hypercharge gauge field relative to the coupling with the Yang-Mills gauge fields.
  Thus observation of the $Z\gamma$ mode as well as three-body and
  four-body decay modes of the $Z'$ will  provide
  a definite test of the model and of a possible  new source of interaction beyond the standard model.

\end{abstract}

\newpage

\section{Introduction}\label{sec:intro}

The ATLAS collaboration at CERN \cite{Aad:2015owa} has seen a diboson excess around $2$~TeV in the $WZ$,
$WW$, and $ZZ$ channels with local significance of 3.4$\sigma$, 2.6$\sigma$, and 2.9$\sigma$ in that order.
In this work we discuss a model where the source of the diboson excess is a $Z'$ boson which gains mass through
the Stueckelberg mechanism~\cite{Kors:2004dx,Cheung:2007ut,Feldman:2007nf,Liu:2011di,Feng:2012jn,Feng:2014eja}
and has interactions with $SU(2)_L$ Yang-Mills gauge bosons.
Our model differs in significant ways from  a
variety of other models that have been proposed to explain the excess.
These include models with strong dynamics~\cite{Fukano:2015hga},
$W^\prime$ models~\cite{Grojean:2011vu}, models based on strings~\cite{Anchordoqui:2015uea},
composite spin zero boson models~\cite{Chiang:2015lqa},
and many others \cite{diboson}.

\section{$Z'$ to Diboson decays}

We consider a $U(1)_X$ extension of the
standard model with $C_{\mu}(x)$ as the gauge boson and we propose
the following effective interaction
\begin{align}
{\cal L} ={ {1 \over  \Lambda^2}} \Big[\Big({1\over M} \partial_\mu \sigma + C_\mu\Big)
\partial_\nu\, \big(F_a^{\mu\lambda}  {F^{~\nu}_{a~\lambda}} + \alpha B^{\mu\lambda} B^\nu_\lambda \big) +
\big(\mu \leftrightarrow \nu\big) \Big]\,,
\label{eq:totalD}
\end{align}
where $F$ and $B$ are the $SU(2)_L$ and $U(1)_Y$ field strengths,
$a$ is the $SU(2)_L$ index, $\alpha$
is the  strength of the coupling of $C_\mu$ with the hypercharge gauge field
relative to the coupling with the Yang-Mills gauge fields and is a free parameter,
and the new physics scale $\Lambda$ will be determined by experiment.
In~\cref{eq:totalD} we are using the Stueckelberg mechanism to make the Lagrangian gauge invariant.
This occurs due to the following gauge transformations of $C_\mu$ and $\sigma$: $C_\mu \to C_\mu -\partial_\mu \lambda\,,\ \sigma \to M \lambda$.
Thus the interactions is  $U(1)_X$ and $SU(2)_L$ gauge invariant.
For more details of  Stueckelberg mechanism and the Stueckelberg $U(1)$ extension of standard model or minimal supersymmetric standard model,  see~\cite{Kors:2004dx}.

In the unitary gauge we define $Z'_\mu = ({1\over M} \partial_\mu \sigma + C_\mu)$.
These interactions will describe the possible decays
of the $Z'$ into two-body, three-body and four-body states.
After expansion, the three-point interactions read
\begin{equation}
\mathcal{L}_{{\rm 3pt}} = \frac{2}{\Lambda^2} \,Z_\mu' \big(
\partial^\nu A^\mu_a \,\partial^2 A_{a\nu}
-\partial^\mu A^\nu_a \,\partial^2 A_{a\nu}
+\alpha \,\partial^\nu B^\mu \,\partial^2 B_{\nu}
-\alpha \,\partial^\mu B^\nu \,\partial^2 B_{\nu} \big)\,.
\label{total2}
\end{equation}
Further after spontaneous breaking of the electroweak symmetry we will have a Lagrangian describing
the interactions  of the $Z'$ with $W, Z$ and $\gamma$.
For the two-body decays the possible modes are
$W^+W^-$, $ZZ$, $Z \gamma$.
In addition, \cref{eq:totalD} also provides three- and four-body decays such as
$W^+W^-Z$, $W^+W^-\gamma$ and $W^+W^-ZZ$ etc.
For the two-body decay, \cref{total2} further reduces to
\begin{align}
\mathcal{L}_{{\rm 3pt}} & =\frac{{2} m_{W}^{2}}{\Lambda^{2}}Z'_{\mu}(\partial^{\nu}W^{-\mu}W_{\nu}^{+}+\partial^{\nu}W^{+\mu}W_{\nu}^{-})\nonumber \\
 & \,+\frac{{2}m_{Z}^{2}}{\Lambda^{2}}Z'_{\mu}\left[
\cos^{2}\theta_{W}
 \partial^{\nu}Z^{\mu}Z_{\nu}
 +\sin\theta_{W}\cos\theta_{W}
 (\partial^{\nu}A^{\mu}Z_{\nu} - \partial^{\mu}A^{\nu}Z_{\nu}) \right] \nonumber \\
  & \,+\frac{{2} m_{Z}^{2}}{\Lambda^{2}}\,\alpha \, Z'_{\mu}\left[
\sin^{2}\theta_{W}
 \partial^{\nu}Z^{\mu}Z_{\nu}
 - \sin\theta_{W}\cos\theta_{W}
 (\partial^{\nu}A^{\mu}Z_{\nu} - \partial^{\mu}A^{\nu}Z_{\nu}) \right] \,,
 \label{3ptL}
\end{align}
where $\sin\theta_{W}=g/ \sqrt{g^{2}+g'^{2}},\cos\theta_{W}=g'/\sqrt{g^{2}+g'^{2}}$.
First we notice $Z'$ to diphoton channel automatically vanishes consistent with the Landau-Yang theorem~\cite{Landau:1948kw,Yang:1950rg,Keung:2008ve,Gninenko:2011ws,Cacciari:2015ela}.
As seen from~\cref{3ptL}
the non-vanishing two-body decays consist of the
final states $W^+W^-$, $ZZ$, $Z\gamma$.
For the case $\alpha = 1$ the $Z'\to Z\gamma$ mode vanishes.
In the analysis of these final states we will use the helicity formalism.
In this formalism the three-point amplitudes for these processes read
\begin{align}
\mathscr{A}&\big[Z'(\xi^{\prime},k')\to W^{+}(\xi_{1}^{+},k_{1})W^{-}(\xi_{2}^{-},k_{2})\big]
=\frac{{2}m_{W}^{2}}{\Lambda^{2}}\big[(\xi^{\prime*}\cdot\xi_{2}^{-})(\xi_{1}^{+}\cdot k_{2})+(\xi^{\prime*}\cdot\xi_{1}^{+})(\xi_{2}^{-}\cdot k_{1})\big]\,,\label{ZptoWW}\\
\mathscr{A}&\big[Z'(\xi^{\prime},k')\to Z(\xi_{1},k_{1})Z(\xi_{2},k_{2})\big]
=\frac{{2}m_{Z}^{2}}{\Lambda^{2}}(\cos^2\theta_{W}+\alpha \sin^2\theta_{W})\times\nonumber\\
&\qquad\qquad\qquad\qquad\qquad\qquad\qquad\qquad\qquad\qquad\qquad
\big[(\xi^{\prime*}\cdot\xi_{2})(\xi_{1}\cdot k_{2})+(\xi^{\prime*}\cdot\xi_{1})(\xi_{2}\cdot k_{1})\big]\,,\label{ZptoZZ}\\
\mathscr{A}&\big[Z'(\xi^{\prime},k')\to Z(\xi,k)\gamma(\epsilon,k_{0})\big]
=\frac{2 m_{Z}^{2}}{\Lambda^{2}} (1-\alpha)\sin\theta_{W}\cos\theta_{W}
\big[(\xi^{\prime*}\cdot\epsilon)(\xi\cdot k_{0})
 - (\xi^{\prime*}\cdot k_0)(\xi\cdot \epsilon)
\big]\,,
\label{2}
\end{align}
where $\xi^{\prime},\xi_{\pm},\xi,\epsilon$ denote the polarization
vector of $Z',W^{\pm},Z,\gamma$.
To see the helicity structure of the above amplitudes more explicitly,
we now apply the spinor helicity formalism. A massless spin one gauge
boson has two degrees of freedom, corresponding to up and
down helicities.
They are expressed by the polarization vectors $\text{\ensuremath{\epsilon}}_{\mu}^{+}$
and $\text{\ensuremath{\epsilon}}_{\mu}^{-}$ and can be written as~\cite{Dixon:1996wi}
\begin{equation}
\text{\ensuremath{\epsilon}}_{\mu}^{+}(k,r)=\frac{r_{\dot{a}}^{*}\bar{\sigma}_{\mu}^{\dot{a}a}k_{a}}{\sqrt{2}r_{\dot{a}}^{*}k^{*\dot{a}}}\,,
\qquad\text{\ensuremath{\epsilon}}_{\mu}^{-}(k,r)=\frac{k_{\dot{a}}^{*}\bar{\sigma}_{\mu}^{\dot{a}a}r_{a}}{\sqrt{2}k^{a}r_{a}}\,,
\end{equation}
where $k$ is the momentum of the particle and $r$ is the reference
momentum which can be chosen to be any light-like momentum except $k$.
Here the momenta with spinor indices are $2$-component commutative spinors,
and they are defined as $p_\mu \sigma^\mu_{a \dot{a}} = - p_a p^*_{\dot{a}}$.
It's easy to show $\epsilon_+^\mu (k,r) -\epsilon_+^\mu (k,\tilde{r}) \sim k^\mu $
where $\tilde{r}$ is some other free chosen reference momentum.
Since the whole amplitude is invariant under the gauge transformation $\epsilon^\mu \to \epsilon^\mu + \lambda k^\mu$,
choosing different reference momentum for a massless gauge boson
does not change the result.

A massive spin one gauge boson
which is expressed by its polarization vector $\xi_\mu$,
contains three degrees of freedom associated
to the eigenstates of $J_{z}$, where the transversality condition $\xi_{\mu}k^{\mu}=0$
eliminates one degree of freedom of the four-vector. The choice of
the quantization axis $\vec{z}$ can be handled in an elegant way
by decomposing the momentum $k^{\mu}$ into two arbitrary light-like
reference momenta $p$ and $q$:
\begin{equation}
k^{\mu}=p^{\mu}+q^{\mu}\,,\qquad k^{2}=-m^{2}=2pq\,,\qquad p^{2}=q^{2}=0\,.
\end{equation}
Once the reference momenta $p$ and $q$ are chosen,
the spin quantization axis of the polarization vector $\xi_\mu$
is set to be collinear to the direction of $\vec{q}$ in the rest frame.
The 3 spin wave functions {depend on} $p$ and $q$,
while this dependence would drop out in the squared amplitudes summing
over all spin directions. The massive spin one wave functions $\xi_{\mu}$
are given by the following polarization vectors (up to a phase factor)~\cite{Spehler:1991yw,Novaes:1991ft}
\begin{align}
\text{\ensuremath{\xi}}^{\mu}(k,J_{z}=+1) & =\frac{1}{\sqrt{2}m}p_{\dot{a}}^{*}\bar{\sigma}^{\mu\dot{a}a}q_{a}\,,\\
\text{\ensuremath{\xi}}^{\mu}(k,J_{z}=\ 0\ ) & =\frac{1}{2m}\bar{\sigma}^{\mu\dot{a}a}(p_{\dot{a}}^{*}p_{a}-q_{\dot{a}}^{*}q_{a})\,,\\
\text{\ensuremath{\xi}}^{\mu}(k,J_{z}=-1) & =-\frac{1}{\sqrt{2}m}q_{\dot{a}}^{*}\bar{\sigma}^{\mu\dot{a}a}p_{a}\,.
\end{align}
With different choices of reference momenta $(p,q)$,
one will get different helicity amplitudes.
While the $(p,q)$-dependence would drop off when one adds up all the squared helicity amplitudes.
Since we have the freedom to choose reference momenta for the interacting
spin one gauge bosons, with a clever choice one can not only simplify
the computation dramatically but also exhibit the helicity structure in a transparent manner.

For the process $Z'(k')\to Z(k)\gamma(k_{0})$,
since the spin quantization axis of a massless photon is collinear to its moving direction,
we choose the following reference momenta
\begin{align}
\gamma \ \qquad \epsilon(k_{0}):&  \qquad r\,,\label{ZGaR1}\\
Z \ \ \qquad \xi(k): & \qquad k=r+ak_{0}\,,\label{ZGaR2}\\
Z' \qquad \xi'(k'):&\qquad k'=r+(a+1)k_{0}\,,\label{ZGaR3}
\end{align}
where $r^{2}=0$ and $a=m_{Z}^{2}/(M_{Z'}^{2}-m_{Z}^{2})$.
For this clever choice, the spin quantization axes of both $Z'$ and $Z$
are aligned to the photon moving direction, i.e., the direction of $\vec{k}_0$.

For  $Z'$ decay into two massive gauge bosons with a common
mass $m_{i}$, where $m_{i}=m_{W}$ for Eq.~\eqref{ZptoWW} and $m_{i}=m_{Z}$
for Eq.~\eqref{ZptoZZ}, we choose the following reference momenta
\begin{align}
Z_1/W^+ \qquad \xi_{1}(k_{1}):&\qquad k_{1}=cp+q\,,\label{2MR1}\\
Z_2/W^- \qquad \xi_{2}(k_{2}):&\qquad k_{2}=p+cq\,,\label{2MR2}\\
Z' \quad\ \ \qquad \xi^{'}(k'):&\qquad k'=(1+c)p+(1+c)q\,,\label{2MR3}
\end{align}
with $m_{i}^{2}=k_{1}^{2}=k_{2}^{2}=2cp\cdot q$, $M_{Z'}^{2}=k'^{2}=2(1+c)^{2}p\cdot q$,
and thus
\begin{equation}
c_i=\frac{1}{2}(b_i-2)\pm\frac{1}{2}\sqrt{b_i^{2}-4b_i}\,,\label{2MRMc}
\end{equation}
where $b_i=M_{Z'}^{2}/m_{i}^{2}$.
Under this choice of reference momenta, the spin quantization axes of all these three massive gauge bosons are aligned to the same direction, i.e., the direction of $\vec{q}$.

\begin{figure}[t!]
\begin{center}
\includegraphics[scale=0.5]{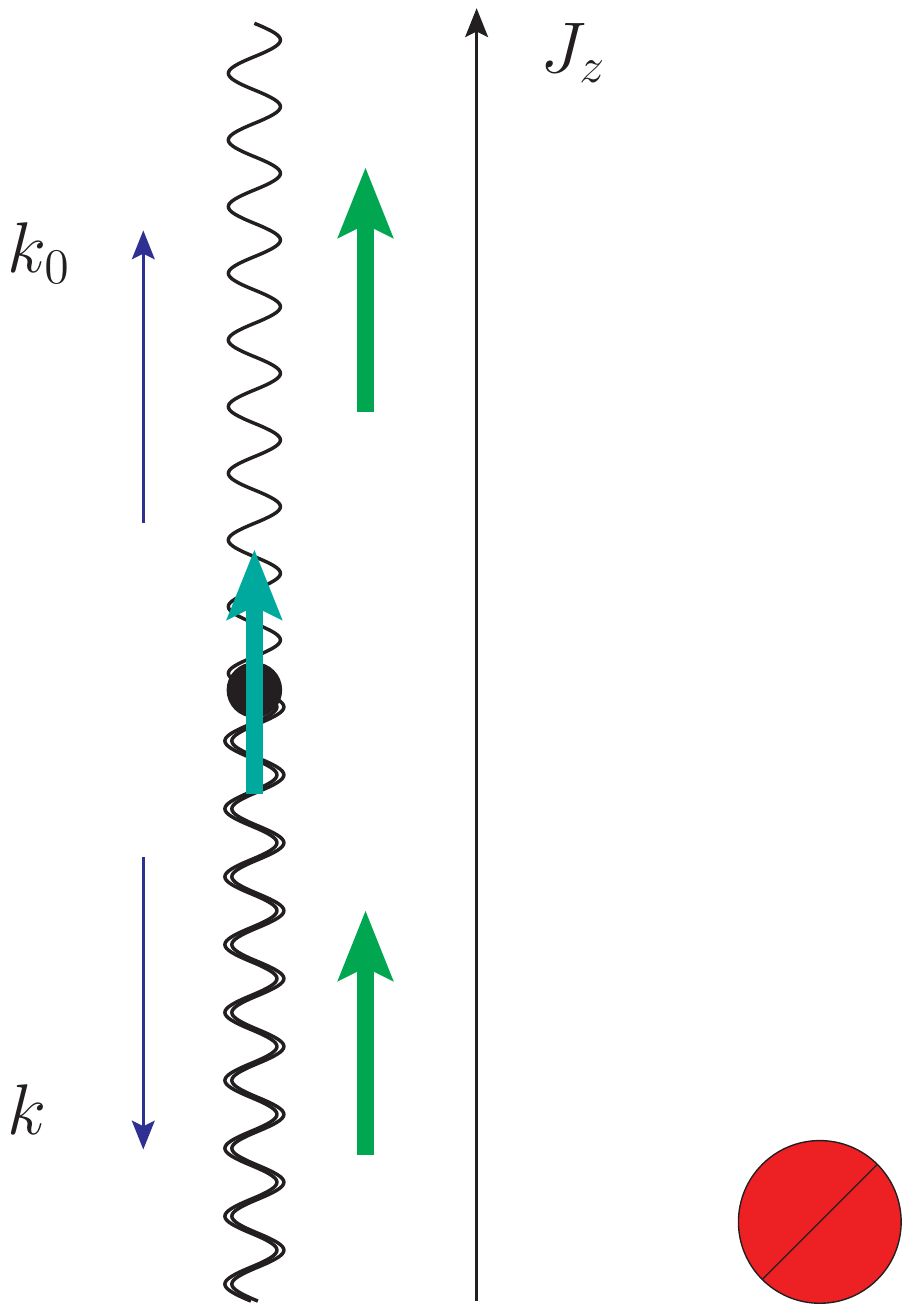}\qquad\qquad\qquad
\includegraphics[scale=0.5]{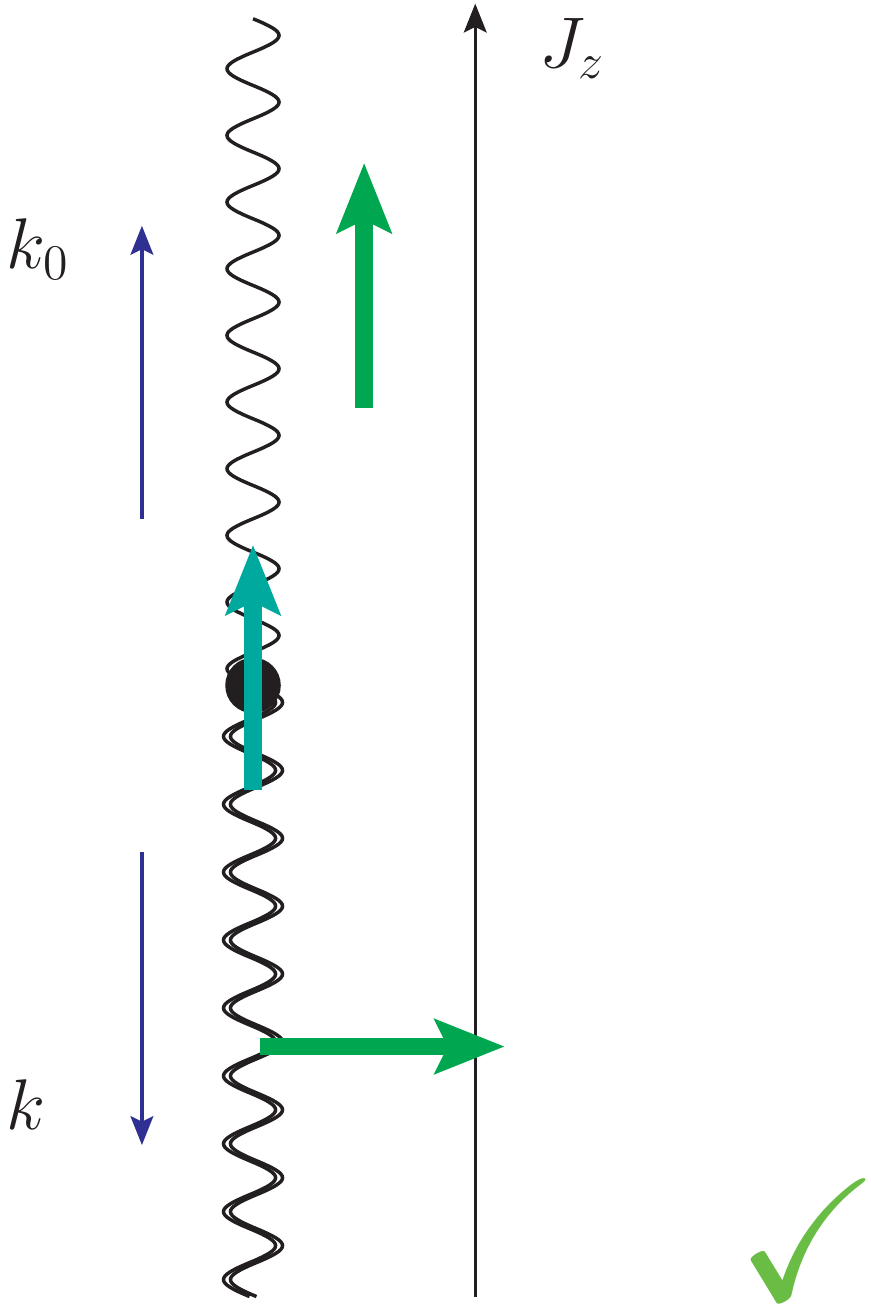}
\caption{[Color online]
The black dots in the center denote the decaying $Z'$.
In the center of mass frame, after decay the photon and $Z$ in the final state are moving to the opposite directions.
The single wavy lines present the photon and the double wavy lines present the $Z$.
The spin of the photon is along (or opposite to) its moving direction.
We choose the reference momenta of the massive $Z$ as well as the $Z'$ in such a way
that their spin quantization axes ($\vec{J_z}$) are aligned to the photon moving direction.
The emerald arrows present the spin of the $Z'$,
and the green arrows show the spin of the photon and $Z$.
The left panel presents the vanishing helicity amplitude
$\mathscr{A}\big[Z' (+) ;Z(+),\gamma(+)\big]=0$
and this process is {\it not allowed} as indicated by the red forbidden sign.
The right panel shows the non-vanishing helicity amplitude $\mathscr{A}\big[Z'(+);Z(0),\gamma(+)\big]$
and this process is {\it allowed} as  indicated by the green check sign.}
\label{FigZpZGa}
\end{center}
\end{figure}

In sum, for the two cases discussed above the spin quantization axes of the decaying massive gauge boson $Z'$ as well as the two gauge bosons in the final state, are aligned to the same direction. Thus for  the process $Z'\to Z\gamma$, it is not difficult to show the vanishing of the following helicity amplitudes
\begin{gather}
\mathscr{A}\big[Z';Z(+),\gamma(+)\big]=\mathscr{A}\big[Z';Z(-),\gamma(-)\big]=0\,, \label{ZGppmm} \\
\mathscr{A}\big[Z'(\pm);Z(+),\gamma(-)\big]=\mathscr{A}\big[Z'(\pm);Z(-),\gamma(+)\big]=0\,, \label{ZGpmmp}
\end{gather}
as a result of angular momentum conservation, c.f., the left panel of~Fig.~\ref{FigZpZGa}.
While the non-vanishing helicity amplitudes are
\begin{align}
\mathscr{A}_{1}&\equiv\mathscr{A}\big[Z'(+);Z(0),\gamma(+)\big]=\mathscr{A}\big[Z'(-);Z(0),\gamma(-)\big]=
\frac{\sin\theta_{W}\cos\theta_{W}}{{\Lambda^{2}}}\beta m_{Z}(M_{Z'}^{2}-m_{Z}^{2})\,, \label{ZGA1} \\
\mathscr{A}_{2}&\equiv\mathscr{A}\big[Z'(0);Z(+),\gamma(-)\big]=\mathscr{A}\big[Z'(0);Z(-),\gamma(+)\big]=
\frac{\sin\theta_{W}\cos\theta_{W}}{{\Lambda^{2}M_{Z'}^2}}\beta m_{Z}^2(M_{Z'}^{2}-m_{Z}^{2})\,,\label{ZGA2}
\end{align}
where $\beta \equiv 1-\alpha$. Thus the total squared-amplitude read
\begin{equation}
Z'\to Z\gamma:\quad\sum|\mathscr{A}|^{2}=2|\mathscr{A}_{1}|^{2}+2|\mathscr{A}_{2}|^{2}
=\frac{{2}\sin^{2}\theta_{W}\cos^{2}\theta_{W}\beta^2}{\Lambda^{4}}\frac{m_{Z}^{2}}{M_{Z'}^2}(M_{Z'}^2+m_{Z}^{2})(M_{Z'}^{2}-m_{Z}^{2})^{2}\,.
\label{pzp-mzm}
\end{equation}
The factor 2 in $2|\mathscr{A}_{1}|^{2}$ and $2|\mathscr{A}_{2}|^{2}$ arise because for each there are two helicity configurations, c.f.,~\cref{ZGA1,ZGA2},
that are non-vanishing.

\begin{figure}[t!]
\begin{center}
\includegraphics[scale=0.5]{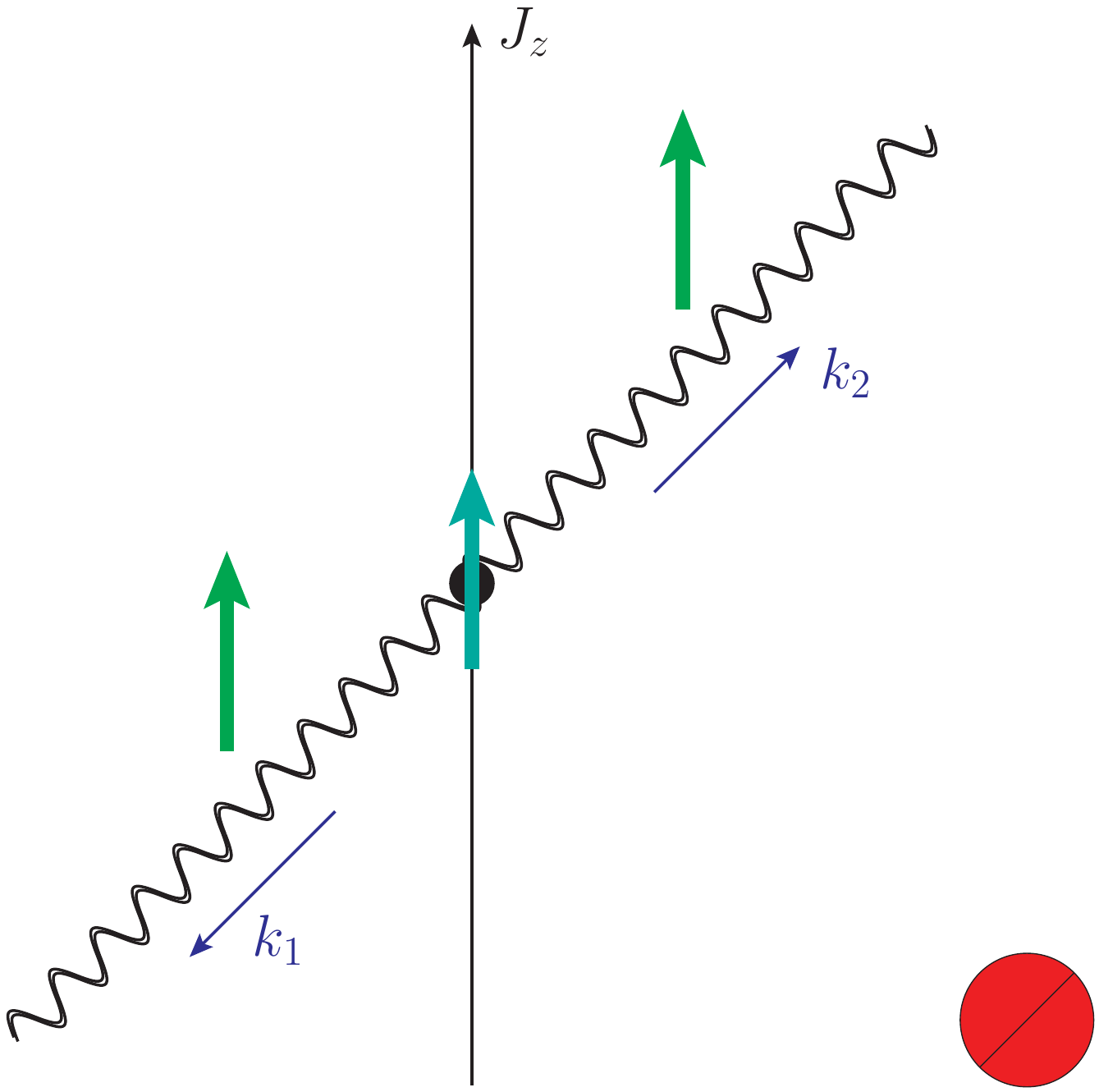}\qquad\qquad
\includegraphics[scale=0.5]{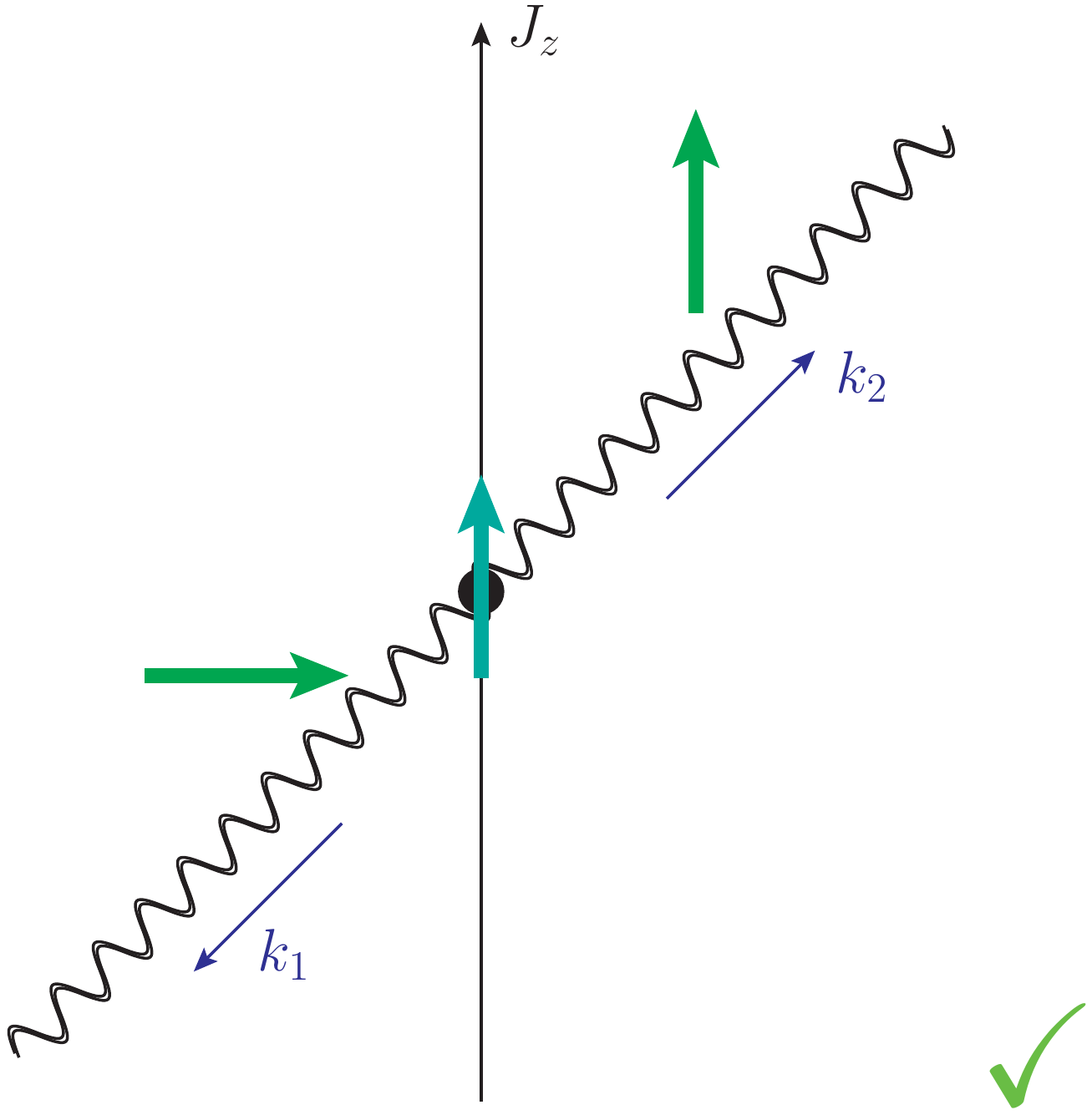}
\caption{[Color online]
In the center of mass frame, after the decay the two massive gauge bosons in the final state are moving to the opposite directions.
The black dots in the center again denote the decaying $Z'$.
We have chosen the reference momenta in such a way that the spin quantization axes of these three gauge bosons are aligned to the same direction, i.e., $\vec{J_z}$.
In general, for a massive gauge boson,
the direction of $\vec{J_z}$ is different with its moving direction
because one can always boost it to other reference frames.
The emerald arrows present the spin of the $Z'$,
and the green arrows show the spin of the two final state gauge bosons.
The left panel presents the vanishing helicity amplitude $\mathscr{A}\big[Z'(+);+,+\big]$,
and the right panel shows the non-vanishing helicity amplitude $\mathscr{A}\big[Z'(+);0,+\big]$.}
\label{FigZp2M}
\end{center}
\end{figure}

Next we discuss the $Z'$ decay into two massive gauge bosons. An analysis similar to the above gives
\begin{gather}
\mathscr{A}\big[Z'(+);+,+\big]=\mathscr{A}\big[Z'(+);-,-\big]=\mathscr{A}\big[Z'(+);0,0\big]=\mathscr{A}\big[Z'(+);+,-\big]=0\,, \label{MM1}\\
\mathscr{A}\big[Z'(-);-,-\big]=\mathscr{A}\big[Z'(-);+,+\big]=\mathscr{A}\big[Z'(-);0,0\big]=\mathscr{A}\big[Z'(-);+,-\big]=0\,,\label{MM2}\\
\mathscr{A}\big[Z'(0);+,+\big]=\mathscr{A}\big[Z'(0);-,-\big]=\mathscr{A}\big[Z'(0);+,0\big]=\mathscr{A}\big[Z'(0);-,0\big]=0\,,\label{MM3}\\
\mathscr{A}\big[Z'(0);+,-\big]=\mathscr{A}\big[Z'(0);-,+\big]=\mathscr{A}\big[Z'(0);0,0\big]=0\,,\label{MM4}
\end{gather}
where the first three lines are due to angular momentum conservation, c.f., the left panel of~Fig.~\ref{FigZp2M}
and the last line is due to the special choice of the reference momenta~\cref{2MR1,2MR2,2MR3}.
The residual set of non-vanishing helicity amplitudes  are
\begin{equation}
\mathscr{A}_{3}\equiv\mathscr{A}\big[Z'(+);+,0\big]=\mathscr{A}\big[Z'(+);0,+\big]=\mathscr{A}\big[Z'(-);-,0\big]=\mathscr{A}\big[Z'(-);0,-\big]\,.\label{Zptomassive}
\end{equation}
Explicitly they are given by
\begin{align}
\mathscr{A}_{3W}&\equiv
\mathscr{A}(Z'\to W^+W^-)
=\frac{m_{W}^{4}}{{\Lambda^{2}}M_{Z'}} \frac{(c_W+1)^{2}(c_W-1)}{c_W^{3/2}}\,, \label{ZptoWWh}\\
\mathscr{A}_{3Z}&\equiv
\mathscr{A}(Z'\to Z Z)
=\frac{(\cos^2\theta_{W}+\alpha \sin^2\theta_{W}) m_{Z}^{4}}{\Lambda^{2}M_{Z'}} \frac{(c_Z+1)^{2}(c_Z-1)}{c_Z^{3/2}}\,,\label{ZptoZZh}
\end{align}
where the coefficients $c_i$ are given in~\cref{2MRMc},
and $m_{i}=m_{W}$ for Eq.~\eqref{ZptoWWh} and $m_{i}=m_{Z}$ for Eq.~\eqref{ZptoZZh}.
The total squared-amplitudes read
\begin{align}
Z'\to W^{+}W^{-}:\qquad & \sum|\mathscr{A}|^{2}=4|\mathscr{A}_{3W}|^{2}
=\frac{{4}}{\Lambda^{4}} M_{Z'}^2 m_{W}^{2}(M_{Z'}^{2}-4m_{W}^{2})\,,\\
Z'\to ZZ:\qquad & \sum|\mathscr{A}|^{2}=2|\mathscr{A}_{3Z}|^{2}
=\frac{2(\cos^2\theta_{W}+\alpha \sin^2\theta_{W})^2}{{\Lambda^{4}}} M_{Z'}^2 m_{Z}^{2}(M_{Z'}^{2}-4m_{Z}^{2})\,.
\end{align}
Here the factor 4 for $W^+W^-$ channel is due to the fact that  there are in total 4 non-vanishing helicity amplitudes in this channel, c.f.,~\cref{Zptomassive}.
The factor 2 for $ZZ$ channel is  due to the fact that there are only 2 non-vanishing helicity amplitudes since the two final state particles are identical,
e.g., $\mathscr{A}\big[Z'(+);+,0\big]$ and $\mathscr{A}\big[Z'(+);0,+\big]$ give the same amplitude for $Z'\to ZZ$ channel.

In summary, by using the helicity formalism, we can see clearly
which helicity modes are forbidden as a result of angular momentum conservation, c.f.,~\cref{ZGppmm,ZGpmmp,MM1,MM2,MM3}
and also~Figs.~\ref{FigZpZGa} and~\ref{FigZp2M}.
In addition, there are multiple helicity amplitudes which
vanish due to the clever choice of the reference momenta, c.f.,~\cref{MM4}.

Using the above the  partial decay widths for the processes $WW$, $ZZ$, $Z\gamma$  are given by
\begin{align}
\Gamma(Z'\to W^{+}W^{-}) &= \frac{1}{{4} \pi} \frac{m_W^2 (M_{Z'}^2-4m_W^2)^{3/2}}{\Lambda^4}\,,\\
\Gamma(Z'\to ZZ) &= \frac{(\cos^2\theta_{W}+\alpha \sin^2\theta_{W})^2}{8 \pi} \frac{m_Z^2 (M_{Z'}^2-4m_Z^2)^{3/2}}{\Lambda^4}\,,\\
\Gamma(Z'\to Z\gamma) &= \frac{(1-\alpha)^2\sin^2\theta_W \cos^2\theta_{W}}{{8} \pi} \frac{m_Z^2 (M_{Z'}^2+m_Z^2) (M_{Z'}^2-m_Z^2)^3}{\Lambda^4 M_{Z'}^5}\,.
\end{align}
In the limit $M^2_{Z'} \gg m^2_W, m^2_Z$ which holds to better than 1\% accuracy one has the following
ratio among the three decay modes
\begin{equation}
\Gamma(W^+W^-): \Gamma(ZZ): \Gamma(Z\gamma)\simeq 2: \cos^2\theta_{W}(1+\alpha \tan^2\theta_{W})^2 : (1-\alpha)^2 \sin^2\theta_W  \,.
\label{RT1}
\end{equation}
For the case $\alpha = 0$, the above ratio reduces to
\begin{equation}
\Gamma(W^+W^-): \Gamma(ZZ): \Gamma(Z\gamma)\simeq 2: \cos^2\theta_W : \sin^2\theta_W \, ,
\label{RT2}
\end{equation}
and for  the case $\alpha = 1$, the above ratio gives
\begin{equation}
\Gamma(W^+W^-): \Gamma(ZZ): \Gamma(Z\gamma)\simeq 2: \cos^{-2}\theta_W : 0 \,.
\label{RT3}
\end{equation}
Thus we see that both the $ZZ$ and the $Z\gamma$ modes are highly model dependent and they could be vanishing or
non-vanishing depending on the value of $\alpha$ (which can be either positive or negative);
more LHC data are needed to fully discriminate the three diboson channels
 and to fix the $\alpha$ parameter.

\section{Phenomenology}

Regarding the coupling of the $Z'$ to the standard model fermions,
we  will assume a leptophobic $Z'$ with the following direct interaction to quarks
\begin{equation}
{\cal L}_\text{int} = g_X Z'_\mu \bar{q} \gamma^\mu q\,.
\end{equation}
The decay width to quarks due to the direct couplings is given by
\begin{equation}
\Gamma_\text{direct} = N_c N_f g_X^2 {M_{Z'} \over 12 \pi}\,,
\end{equation}
where $N_c=3$ is the QCD color factor, and $N_f$ is the number of quark flavors
that the $Z'$ can decay into which is the number of  kinematically allowed flavors.
Without going into details we assume that our $U(1)_X$ with a gauged  baryon number
is anomaly free. Such a $U(1)_X$ can arise in a  variety of settings such as
from gut models~\cite{Babu:1996vt},
or anomaly-free family-dependent $U(1)$'s~\cite{Crivellin:2015lwa},
with extra heavy chiral particles to cancel the anomaly~\cite{FileviezPerez:2010gw,Dulaney:2010dj,FileviezPerez:2011pt,Duerr:2013dza}.
We further assume that the heavy chiral states are not accessible at the current LHC  energy and thus do not enter in $Z'$ decay.\\

We discuss now the  production cross section of the $Z'$ at LHC at $\sqrt{s}=8$ TeV and estimate  the size of the diboson excess.
The parton level cross section for the process $q\bar q \to Z' \to W^+W^-$ using
Breit-Wigner form for the $Z'$ intermediate state is given by
\begin{align}
\hat{\sigma} (q\bar q \to Z' \to W^+ W^-) =
{1 \over 12 \pi \Lambda_{\rm eff}^4}\sqrt{1-4 m_W^2/\hat{s}}
{\hat{s} M_{Z'}^2 (\hat{s} - 4 M_{Z'}^2 x_W^2) x_W^2 \over
(\hat{s}-M_{Z'}^2)^2 + \hat{s}^2 \Gamma_{Z'}^2 M_{Z'}^{-2}}\,,
\label{z.1}
\end{align}
where $\Lambda^2_\text{eff} = \Lambda^2/g_X$,
$x_W\equiv m_W/m_{Z'}$.
The $q\bar q \to Z' \to ZZ$ cross section can be obtained easily if
one replaces $m_W$ by $m_Z$ and inserts the overall factor
{$(\cos^2\theta_W+\alpha \sin^2\theta_W)^2/2$}
to the above $W^+W^-$ cross section.
The hadron cross section at  the LHC ($\sqrt{s}=8$ TeV) is computed via
the convolution
\begin{align}
\sigma(pp\to Z'\to W^+W^-) = K \int d\tau\, \hat{\sigma}(\hat{s}=s \tau)\,
{d{\cal L} \over d \tau} (\tau)
\label{eq:LHCxsec}
\end{align}
where we use $K \simeq 1.3$ \cite{Accomando:2010fz} \cite{Cao:2012ng}
to approximate the next to leading correction and
${d{\cal L}/d \tau}$ is the parton luminosity given by
${d{\cal L}/ d \tau} (\tau) = 2 \int_\tau^1 (dx/x)
\left[ u(x) \bar{u}\left({\tau/x}\right)+d(x) \bar{d} \left({\tau/x}\right) \right]$.

\begin{figure}[t!]
\vspace{0.4cm}
\centering
\includegraphics[width=0.48\textwidth]{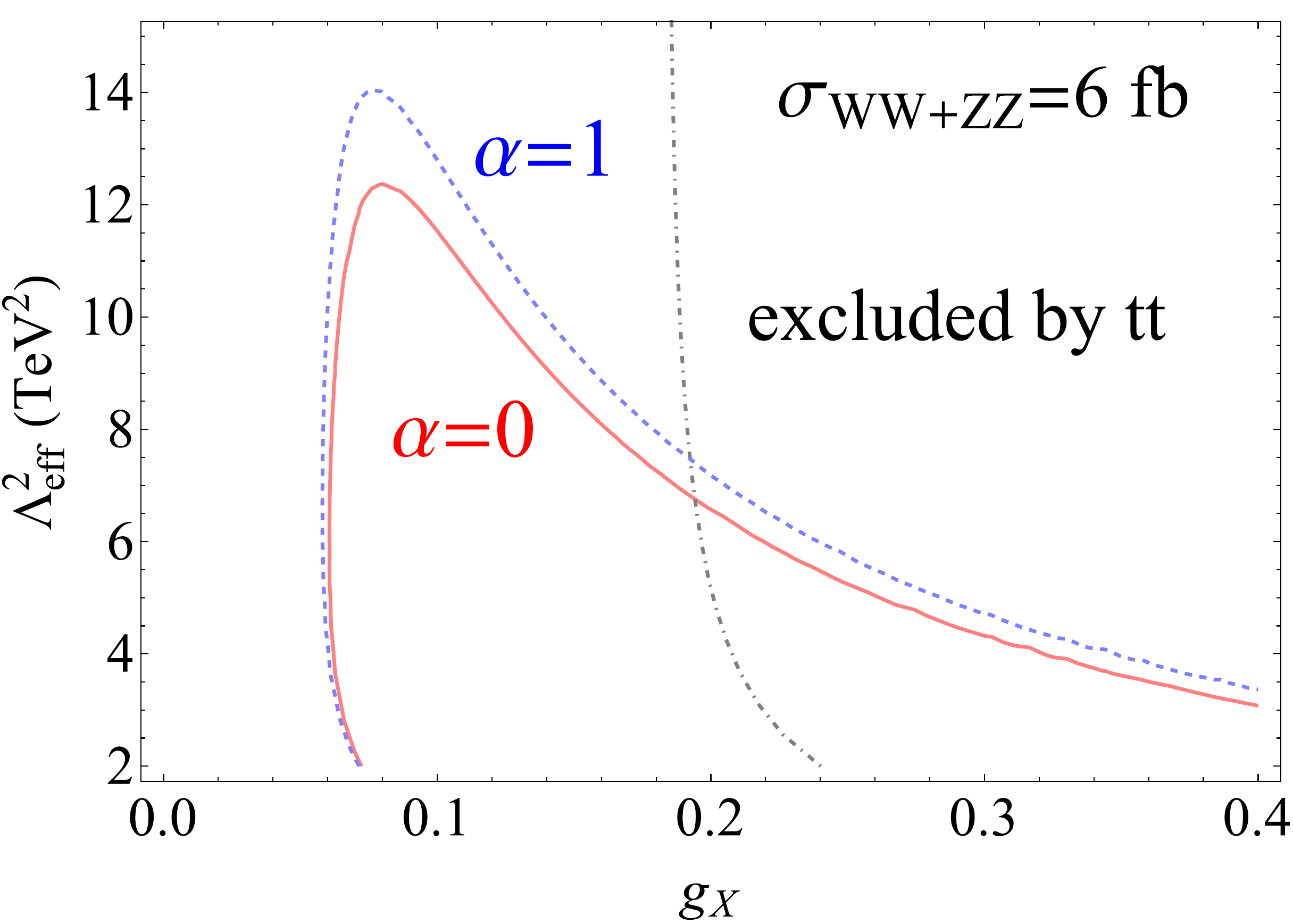}
\hspace{0.2cm}
\includegraphics[width=0.45\textwidth]{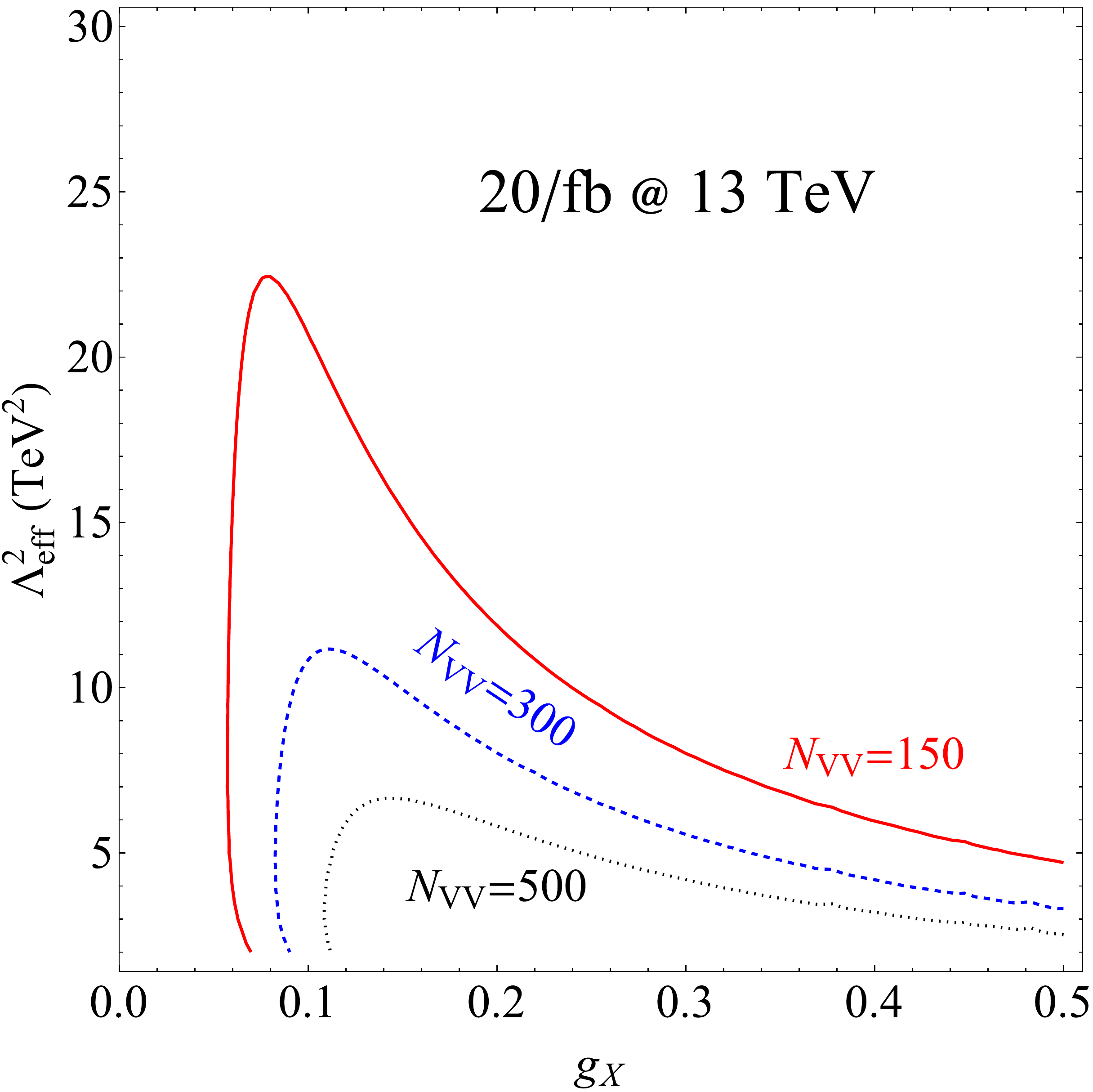}
\caption{Left panel: Analysis at $\sqrt s = 8$ TeV using events over the energy range
  $1.8$ TeV $<\sqrt{\hat{s}}<2.2$ TeV computed for both $\alpha=0$ and $\alpha=1$ cases:
  The red solid (blue dashed) curve predicts $\sigma_{\rm WW+ZZ}=6$ fb
  for the $\alpha=0$ ($\alpha=1$) case, where
  the Breit-Wigner effects of the $Z'$ resonance have been taken into account.
  The parameter region on the right hand side of the gray dot-dashed curve is excluded by the
  $t\bar{t}$ resonance search at the LHC.
  Right panel: Expected number of events in the energy range $1.8$ TeV $<m_{VV}<2.2$ TeV
  with an integrated luminosity 20 fb$^{-1}$ at LHC run~II at $\sqrt s=13$ TeV.
  Here $VV$ includes both $WW$ and $ZZ$ events.
  We take $\alpha=0$ here,
  and assume the same event selection efficiency with event topology requirements
 as given in the ATLAS analysis  \cite{Aad:2015owa}, which is  $\sim$80\%.}
\label{fig:ww}
\end{figure}

The most stringent LHC constraints for the leptophobic $Z'$ come from the $t\bar{t}$ resonance search
\cite{Khachatryan:2015sma} \cite{ATLAS:2015aka}, and the dijet channel
\cite{Khachatryan:2015sja} \cite{Aad:2014aqa}. The 95\% CL upper limit on
dijet cross section for a 2 TeV $Z'$ is
$\sigma(Z')\times \text{BR}(Z' \to q\bar{q}) < 100$ fb \cite{Khachatryan:2015sja}.
For $t\bar{t}$ resonance search, the 95\% CL upper limit for a 2 TeV $Z'$ is
$\sigma(Z')\times \text{BR}(Z' \to t\bar{t}) < 11$ (18) fb when
$\Gamma_{Z'}=20$ (200) GeV \cite{Khachatryan:2015sma}.
We use 11 fb as the limit in the $t\bar{t}$ channel for $\Gamma_{Z'} \leq 20$ GeV, 18 fb for
$\Gamma_{Z'} \geq 200$ GeV, and linearly interpolate these two values for
decay widths in between. Because the $Z'$ boson couples universally to all quarks
in our model, the current $t\bar{t}$ constraint turns to be almost always stronger than the current dijet constraint at the LHC.
Thus, we only consider the $t\bar{t}$ constraint in our analysis.

An analysis of the diboson excess in the $\Lambda^2_{\rm eff}-g_X$ plane is exhibited in~Fig.~\ref{fig:ww}.
Thus the left panel of~Fig.~\ref{fig:ww} gives the prediction of the model at $\sqrt s=8$ TeV.
The red solid curve gives rise to a diboson cross section
$\sigma_{\rm WW+ ZZ}=6$ fb at LHC for $\sqrt{s}=8$ TeV, where $\alpha=0$ is taken.

For comparison we give the analysis {by taking $\alpha=1$}
which is shown by the blue dashed curve which are shifted upward
relative to the red solid curve, in the left panel of~Fig.~\ref{fig:ww}.
A prediction of what we will see at $\sqrt s=13$ TeV at the LHC run~II is given
in the right panel of~Fig.~\ref{fig:ww} in terms of the expected number of events at an integrated
luminosity of 20 fb$^{-1}$.
The LHC production cross section at 13 TeV of the 2 TeV $Z'$ boson in our model is about
7 times larger than at 8 TeV: $\sigma_\text{13 TeV}/\sigma_\text{8 TeV} \simeq 7$.
In the LHC run~II results recently released, the diboson excess events near 2 TeV observed at the 8 TeV data
are not seen in the new 13 TeV data \cite{ATLAS:diboson:run2} \cite{CMS:2015nmz}.
However, because the new ATLAS (CMS) data consist of  an integrated luminosity of $\sim 3.2 ~(2.6)$ fb$^{-1}$ only,
its discovery potential is not improved compared to the 8 TeV data with 20 fb$^{-1}$ integrated luminosity.
Thus we take an agnostic attitude towards the diboson excess events
and await future LHC data with larger luminosity to sort this anomaly out.
If less excess events are seen in the future data, the parameter $\Lambda_\text{eff}$ should be increased to larger values for a given $g_X$ value.

\section{Conclusion}\label{sec4}

In this work we have investigated the diboson excess seen at ATLAS via the decay of a leptophobic
Stueckelberg $Z'$ boson with a mass around 2 TeV. It is possible to accommodate the diboson excess seen by
the ATLAS collaboration within the model of~\cref{eq:totalD}.  Further, the model makes the prediction of a $Z\gamma$ mode
which should also be seen. Additionally the model predicts three-body decays  such as $WWZ$, $WW\gamma$
and four-body decay modes such as  $WWWW$, $WWZZ$, $WW Z\gamma$ etc. Observation of such modes would
provide a confirmation of the proposed model. We also make estimates of the diboson cross sections at LHC run~II.

The proposed model contains new interactions involving vertices
$Z'ZZ, Z'WW, Z'Z\gamma$ which can contribute to the oblique parameters.
Specifically, corrections to the  parameter $S$ can arise from the  $Z-Z$,  $\gamma\gamma$ and $Z-\gamma$ self-energy diagrams~\cite{peskin}.
From \cref{3ptL}  we see that  the effective coupling is of the size $g_{\rm eff} \sim O(M_Z^2/\Lambda^2)$.
Taking $\Lambda^2 \sim 10~{\rm TeV}^2$, we have that $g_{\rm eff}\sim 10^{-3}$ and the
loop is proportional to $\alpha_{\rm eff}=g_{\rm eff}^2/(4\pi) \sim 10^{-7}$.
This is to be compared to the electroweak fine-structure constant $\alpha_2\sim 0.033$.
Thus the contribution from the new physics loops to the $S$
parameter would be much smaller that the current error corridor on $S$ as can be seen from the
global gfitter~\cite{gfitter} results which give $ S=0.05\pm 0.11$.

The proposed interaction~\cref{eq:totalD} is phenomenological and it should be interesting to look for an ultraviolet complete model
that can give rise to such an interaction if the results of LHC run~I are confirmed in LHC run~II.
 Our purpose in investigating the model of \cref{eq:totalD} is to show that there exists an interaction which could
produce the desired diboson resonance. The largeness of the effect seen demands that the
effective scale $\Lambda$ be not too high. A more fundamental model which replaces \cref{eq:totalD}
would only readjust the parameters but our main hypothesis that any fundamental interaction
that can produce \cref{eq:totalD} can explain the experimental observation would still hold.
 An interesting
attribute of \cref{eq:totalD} is that it is a CP-violating interaction and thus a check of this model and specifically
of \cref{RT1,RT2,RT3}  implies that one is testing a new source of CP violation which is accessible at
LHC energies. We note that $\Lambda$ is not necessarily  the mass of a field but a composite
scale, and the mass of the heavy field that gives rise to \cref{eq:totalD} could be much higher.
Consider, for example, a two-index field $\sigma_{\mu\nu}$ with a Lagrangian interaction
$\mathcal{L} \sim m_1^2 \sigma_{\mu\nu}\sigma^{\mu\nu} + m_2^{-1} \sigma_{\mu\nu} J^{\mu\nu}$
with $J^{\mu\nu}= F^{\mu\lambda}F^\nu_{~\lambda} + m_3^2(\partial^\mu C^\nu + \partial^\nu C^\mu)$.
Integration on the $\sigma$ field leads to the interaction of \cref{eq:totalD} with $\Lambda \sim m_1 m_2/m_3$. It is clear
that the choice $m_3/m_2\simeq 6$ will lead to $m_1\sim 10$ TeV, i.e., the mass of the heavy field
would be significantly higher than the resonance mass.  \\

\noindent
{\it Acknowledgments:}
WZF is grateful to Yang Zhang for helpful discussions.
WZF is supported by the Alexander von Humboldt Foundation and Max--Planck--Institut f\"ur Physik, M\"unchen.
The work of Z.L.\ is supported in part by the Tsinghua University Grant 523081007.
The work of PN is supported in part by the U.S. National Science Foundation (NSF) grant PHY-1314774.

\end{document}